# Microbial Corrosion Prevention by *Citrobacter* sp. Biofilms

Pawan Sigdel[1,3], Ananth Kandadai[2,3], Kalimuthu Jawaharraj[1,3,4], Bharat Jasthi[2,3,4], Etienne Gnimpieba[3,4,5], Venkataramana Gadhamshetty[1,3,4]*

[1]Civil and Environmental Engineering, South Dakota School of Mines and Technology, 501 E. St. Joseph Street, Rapid City, SD, 57701, USA

[2]Materials and Metallurgical Engineering, South Dakota School of Mines and Technology, 501 E. St. Joseph Street, Rapid City, SD, 57701, USA

[3]2D-materials for Biofilm Engineering, Science and Technology (2DBEST) Center, South Dakota School of Mines and Technology, 501 E. St. Joseph Street, Rapid City, SD, 57701, USA

[4]Data-Driven Materials Discovery for Bioengineering Innovation Center, South Dakota Mines, 501 E. St. Joseph Street, Rapid City, SD, 57701, USA

[5]Biomedical Engineering, University of South Dakota, 4800 N Career Ave, Sioux Falls, SD 57107, USA

*Corresponding author.

*E-mail address:* Venkata.Gadhamshetty@sdsmt.edu (V. Gadhamshetty).

**Abstract**

Microbiologically influenced corrosion (MIC) compromises the integrity of many technologically relevant metals. Protective coatings based on synthetic materials pose potential environmental impacts. Here, we report a MIC resistant coating based on a biofilm matrix of *Citrobacter* sp. strain MIC21 on underlying copper (Cu) surfaces. Three identical corrosion cells varying in the type of working electrode (annealed Cu, 29.5% coldworked, and 56.2% coldworked Cu) were used. Graphite plate and Ag/AgCl served as counter and reference electrodes, respectively. The working electrolyte was based on lactate-C media along with an inocula consisting of *Oleidesulfovibrio alaskensis* strain G20 and *Citrobacter* sp. strain MIC21. Passivating effect of the co-cultured biofilm matrix was observed in the form of an ennoblement effect. Tests based on sequencing, microscopy, and spectroscopy revealed the formation of a compact biofilm matrix dominated by strain MIC21 cells, exopolymers, and insoluble precipitates. This matrix displayed elastic modulus (a measure of rigidity) as high as 0.8 Gpa and increased corrosion resistance by ~10-fold. Interestingly, strain MIC21 has the capacity to inhibit the undesirable growth of aggressive strain G20. Additional corrosion tests also substantiated the passivation effects of strain MIC21. We provide mechanistic insight into the underlying reasons responsible for corrosion prevention behavior of the biofilm matrix.

## 1. Introduction

Microbiologically influenced corrosion (MIC), also known as microbial corrosion or biocorrosion, refers to the accelerated degradation of metals in the presence of microorganisms [1]. The US Airforce alone spends $1 billion annually to address MIC effects caused by sulfate reducing bacteria (SRB) alone [2]. Robust metals, including copper alloys that resist oxidation under abiotic conditions [3], tend to fail in microbial environments [4-6]. The reasons for these failures can be understood by reviewing different types of MIC mechanisms (e.g., cathodic depolarization (1934), King's Mechanism (1971), anodic depolarization (1984), Romero mechanism (2005), biocatalytic cathodic sulfate reduction (2009)) [7] (See Table S1 for details on SRB). Any given mechanism will involve a series of redox reactions, for example, a thermodynamic coupling between lactate oxidation (Eqn 1) and sulfate reduction (Eqn 2) [7, 8] in the case of SRB.

$$2CH_3CHOHCOO^- + 2H_2O \rightarrow 2CH_3COO^- + 2CO_2 + 8H^+ + 8e^- \;(E^{o\prime} = -430 \; mV) \quad (1)$$

$$SO_4^{2-} + 9H^+ + 8e^- \rightarrow HS^- + 4H_2O \;(E^{o\prime} = -217 \; mV) \quad (2)$$

where $E^{o\prime}$ = modified reduction potential at pH 7, 1 M solutes, or 1 bar gases at 25 ºC.

The bisulfide (HS⁻) from Eqn 2 combines with the hydrogen ions to generate hydrogen sulfide (H₂S) (Eqn 3).

$$HS^- + H^+ \rightarrow H_2S \quad (3)$$

The cell potential ($\Delta E^{o\prime} = +213 \; mV$) from Eqn (1-2) yields a negative Gibbs free energy change ($\Delta G^{o\prime} = -164 \; kJ/mol$) under standard conditions [9], implying a favorable thermodynamic coupling. *The* ΔG values were determined using the following Eqn [10]:

$$\Delta G^0 = -nFE^0_{rxn} \tag{4}$$

where, $n$ = number of electrons passed per atom, $F$ = charge on a mole of electrons, and $E^0_{rxn}$ = standard electromotive force (emf) of the cell reaction.

Coupling between Cu oxidation (Cu$^+$/Cu; $E^{o\prime} = +520\ mV$; Cu$^{2+}$/Cu; $E^{o\prime} = +340\ mV$ @25 °$C$, pH = 7), and sulfate reduction yields negative cell potential. Based on this scenario, one may interpret that the Cu metals are not vulnerable under ambient and neutral pH conditions. However, the bisulfide generated by the sulfate reducing bacteria (Eqn 2) reacts with Cu ions to generate copper sulfide (Cu$_2$S) ($\Delta G^{o\prime} = -62\ kJ/mol$) [8], promotes copper corrosion under ambient conditions (Eqn 1, 2, 5), which is impossible in the absence of microorganisms.

$$2Cu(crystal) + HS^- + H^+ \rightarrow Cu_2S(crystal) + H_2(g) \tag{5}$$

Despite the antimicrobial properties of Cu [11], SRB cells colonize Cu surfaces by forming biofilms, where they encapsulate themselves within a self-secreted extracellular polymeric substance (EPS) [12]. Such biofilms overcome Cu stress by *(i)* using sulfide metabolites (Eqn 2) to reduce Cu ions; *(ii)* restricting permeation of Cu ions from the outer and inner membranes, cytoplasm, and periplasm; *(iii)* scavenging Cu ions using proteins; and *(iv)* expelling Cu ions from the cells [11].

Protective coatings are typically used to delay the onset of corrosion in both abiotic and microbial environments. Although they can effectively resist the abiotic forms of corrosion, they are not necessarily suitable for controlling the adherence state and biofilm growth of detrimental microorganisms, including SRB cells. To improve fouling properties, such coatings are often modified with biocides (e.g., tributyltin) and antimicrobial particles (e.g., silver nanoparticles).

Owing to their toxic effects and potential environmental impacts when discharged into the ecosystem, these coatings are being banned in many countries [13, 14]. Furthermore, any protective coatings can exert influence only on the first layer of adhered bacteria, with much less influence on the invasion by other colonizing microorganisms. To address these issues, the scientific community is beginning to explore a microbiologically induced corrosion inhibition (MICI) method that involves the use of living microorganisms for corrosion prevention. The MICI effects have been attributed to the factors related to microbial respiration, EPS protection, mineralization, competition, and secretion of corrosion inhibitors [15]. For example, the axenic biofilms of *Pseudomonas fragi* and *Escherichia coli* DH5α have been reported to inhibit steel corrosion by forming a low-oxygen barrier, as reported by Jayaraman and his coworkers [16]. *S.Oneidensis* MR-1, a facultative anaerobe, has also been reported to display passivation effects by creating low-oxygen atmosphere for inducing anaerobic respiration with reduction of Fe (III) to Fe (II) [17].

Drawing inspiration from the presence of *Citrobacter* sp. in both human and animal intestinal tract, we use the term "commensal" to define beneficial bacteria that defend metal surfaces against the colonization of corrosive bacteria. As shown in Table S2, most of the prior studies focused only on using axenic cultures of bacteria for passivating corrosion. Here, we explore the use of *Citrobacter* sp*.,* a Gram-negative bacterium within the *Enterobacteriaceae* family, and the *Gammaproteobacteria* class as a commensal bacterium for defending copper surfaces against the corrosive effects of SRB. *Citrobacter* sp. are generally considered commensal bacteria in a healthy human gut [18]. They have also been reported to be found in marine environments facing ship hulls. Although they co-exist with community members involved in the corrosion of ship hulls [19], they have been reported to play a neutral role towards corrosion.

They thrive in soil, sewage, and water environments [20, 21], as well as in engineered systems [22]. Unlike typical SRB, *Citrobacter* facilitates dissimilatory sulfate reduction under micro-aerobic conditions (see Table S3 for prior studies with *Citrobacter)*.

We explored the passivation behavior of *Citrobacter* sp. strain MIC21 grown on different Cu substrates (annealed Cu, 29.5%, and 56.2% coldworked Cu). A co-culture of *Oleidesulfovibrio alaskensis* strain G20 (strict anaerobe) and *Citrobacter* sp. strain MIC21 (facultative anaerobe) was used as the inocula. *O. alaskensis* strain G20 was the lab-maintained model SRB. *Citrobacter* sp. was isolated from lab-cultivated SRB consortia. Strain MIC21 was found to outcompete strain G20 and evolve into a compact biofilm at the end of the corrosion tests. Results based on Open Circuit Potential (OCP), Electrochemical Impedance Spectroscopy (EIS), and an equivalent electrical circuit fitted with the EIS data were used to quantify the passivating behavior of strain MIC21. The corrosion prevention performance of the biofilm matrix was quantified in terms of corrosion resistance, pore resistance, and charge transfer resistance. To explore the underlying passivation mechanisms of MIC21 biofilm, tests based on 16S rRNA sequencing, Scanning Electron Microscopy (SEM), Energy Dispersive Spectroscopy (EDS), Confocal Laser Scanning Microscopy (CLSM), nanoindentation, and X-Ray Diffraction (XRD) were carried out. Additional MIC tests using individual cultures of *Citrobacter* sp. strain MIC21, and *O. alaskensis* strain G20 were used to further corroborate the passivation behavior of MIC21 compared to G20.

## 2. Materials and methods

### 2.1 Copper samples

Cu cylinders (99.95% purity, 1-inch diameter, Online Metals, USA) were sectioned into 1-inch discs (thickness =1.5 mm), polished manually using 340, 300, 400, and 600 mesh silicon carbide (SiC) papers, and rinsed in acetone to remove any contaminants. The treated samples were annealed at 950 ºC for 1 h in an argon atmosphere and cooled to room temperature at 5 °C/min. Two of the annealed samples were subjected to cold working (CW) to achieve a thickness of 29.5% and 56.2%, respectively. The CW process entailed cold rolling of the annealed copper to achieve the samples with 29.5% and 56.2% of the original thicknesses, respectively. The CW process introduced stresses in the samples. The CW Cu by cold rolling is a common practice of copper manufacturing for marine industries and oil exploration. These samples were rinsed with distilled water and alcohol and air-dried prior to use in corrosion tests. The latter steps ensured consistent elemental composition in all the tested samples.

## 2.2 Cultures of strain G20 and MIC21 for MIC studies

Lactate-C media was used to grow individual cultures of the G20 and MIC21 strains as well as their co-culture. Lactate-C media consisted of sodium lactate (6.8 g/L), sodium sulfate (4.5 g/L), sodium citrate (0.3 g/L), dehydrated calcium chloride (0.06 g/L), ammonium chloride (1.0 g/L), magnesium sulfate (2.0 g/L), potassium phosphate monobasic (0.5 g/L), and yeast extract (1.0 g/L). *Citrobacter* sp. strain MIC21, and *O. alaskensis* strain G20 were grown separately in 150 mL serum bottles consisting of 63 mL media and 7 mL inocula. The introduction of the inocula was preceded by the following steps: (i) adjust the initial pH of the media to 7, (ii) seal the bottle along with the media using a rubber septum and crimp, (iii) purge with $N_2$ (95% v/v) and $H_2$ (5% v/v), respectively, followed by (iv) autoclaving at 121 °C for 20 minutes. The incubation was carried out at 30 °C. The optical density of both cultures was nearly 0.2 prior to the use. The doubling time was also investigated for the individual cultures of strain MIC21 under both aerobic and anaerobic conditions along with individual culture of strain G20 under anaerobic conditions [23].

**2.3 Microbiologically influenced corrosion (MIC) tests**

Three different MIC reactors varying in the type of working electrode, namely annealed Cu, 29.5% CW, and 56.2% CW, were setup. The working electrodes were mounted on the stainless-steel sample brackets (Gamry part number 990/00254), and an electroplating tape was to expose 1 cm$^2$ of the electrode to the electrolyte. The reference electrode and counter electrode was based on Ag/AgCl and graphite plate, respectively. The electrolyte consisted of 360 mL of lactate-C media and 40 mL of the co-culture (i.e., MIC21 and G20 (50% v/v)). Additional corrosion tests were carried out using 40 mL of individual cultures of *Citrobacter* sp. strain MIC21 and *O. alaskensis* strain G20. These tests were based on annealed Cu as the working electrode. The Gas chromatography (SRI Instruments, model 8610C, CA, USA) equipped with a thermal conductivity detector and a molecular sieve column (Restek Mole sieve 5A 80/100 1.83m $\times$ 38mm $\times$ 26 mm) was used to measure the composition of the headspace in the test reactors.

**2.4 Weight loss measurement**

The weight loss measurements were conducted using modified ASTM G 31 [24]. Three sets of immersion tests were carried out using annealed Cu and Lactate-C media. These tests varied in the type of inocula, which included (1) Co-culture of *Citrobacter* sp. strain MIC21 and *O. alaskensis* strain G20 (2) Individual culture of strain MIC21, (3) Individual culture of strain G20. Duplicate test specimens were used in each test. Testing apparatus was based on serum bottles containing 70 mL culture and 80 mL headspace. These bottles were crimped with rubber septum to achieve a tight atmospheric seal. Nitrogen was purged to maintain anaerobic conditions for the test (3). A circular Cu coupon of 12.7 mm diameter and 1 mm thick with a surface area of 10.94 cm$^2$ was used as the test specimen. The specimens were precleaned using acetone and methanol and were air-dried before measuring the initial weights. The serum bottles were maintained at 30 °C under stagnant conditions. At the end of the tests, which lasted for 15 days, the Cu coupons were cleaned using ASTM G1 standards [25] and air-dried before measuring the final weights.

**2.5 Electrochemical measurements**

A Gamry potentiostat (Interface 1010) was used to ascertain a steady-state open circuit potential (OCP) value prior to all the electrochemical impedance spectroscopy (EIS) tests. These tests were performed using a 10 mV AC signal within a frequency range of 100 kHz and 0.01 Hz. The EIS spectra were obtained in the forms of Nyquist and Bode curves, respectively, and fitted to an appropriate electrical equivalent circuit (EEC) to determine the impedance between the working electrode and the reference electrode (Section 3.2). As shown in Figure 3a, $R_{soln}$ is the resistance offered by the electrolyte, $R_{po}$ is the pore resistance of the biofilm matrix, $R_{ct}$ is charge transfer resistance. The constant phase element, $C_{po}$, is the capacitance of porous biofilm. $C_{dl}$ is double-layer capacitance at the underlying Cu surface. The $R_{po}C_{po}$ is pore resistance offered by the biofilm and $R_{ct}C_{dl}$, the corrosion process at the interface of electrode and biofilm. A constant phase element was used instead of capacitance [26] and converted into capacitance as follows [2].

$$C = R^{(\frac{1-n}{n})}.Q^{\frac{1}{n}} \tag{6}$$

Where, C = capacitance

Q = Constant Phase element capacitance as $C_{dl}$ or $C_{po}$

R = Resistance

n = exponent in constant phase element

Potentiodynamic polarization curves were measured with the potential sweep of ±250 mV vs. Ag/AgCl. The measurements were carried out at the end of the tests for those based on the co-culture as well as individual cultures of *Citrobacter* sp. and *O. alaskensis*.

**2.6 Genomic DNA extraction and molecular characterization**

The planktonic cells under the co-culture conditions were harvested from the corrosion cells at the end of the tests. The extracted cells were centrifuged at 8000 rpm for 15 min. The genomic DNA was extracted using a DNA extraction kit following the manufacturer's protocol (PureLink™ Microbiome DNA Purification Kit). Molecular characterization was performed using 16S rRNA gene amplification and gene sequencing techniques. The PCR tests were performed with 100 ng of genomic DNA template using 8F and 1492R as universal primers (Turner et al., 1999). The 16S rRNA conserved region was amplified using 5' – AGAGTTTGATCCTGGCTCAG - 3' and 5' – GGTTACCTTGTTACGACTT – 3' as forward and reverse primers, respectively. A 50 µL reaction was performed using a 2X PCR master mix (Platinum™ II Hot-Start Green PCR Master Mix) along with 4 µL of genomic DNA, 4 µL of forward and reverse primers each (0.4 µM), and 13 µL of RT-PCR grade water (Invitrogen). PCR thermal cycler (MiniAmp Plus, Applied Biosystems) was programmed for the following PCR conditions: initial denaturation at 95 °C for 5 minutes, 35 cycles of denaturation at 95 °C for 1 minute, annealing at 60 °C for 1 minute, extension at 72 °C for 1 minute, and final extension at 72 °C for 10 minutes. Negative control was maintained using RT-PCR grade water as the PCR template. The amplified PCR products were separated using a 1.2% agarose gel electrophoresis for 30 min at 60 V in an electrophoretic chamber. The PCR amplicons were visualized and confirmed with ethidium bromide staining using a UV transilluminator. The positive PCR amplicons at ~1500 bp were excised and eluted using a gel purification kit (PureLink™ Quick

Gel Extraction Kit). The amplicon was sequenced using Sanger's dideoxy gene sequencing methods and subjected to BLAST analysis (www.ncbi.nlm.nih.gov). The homologous sequences were identified and compared using the e value and query coverage from the published 16S rRNA sequences. A phylogenetic tree was constructed using a neighbor-joining tree method of the phylogeny test. Here, 1000 bootstrap replications were used using a MEGA 11 software employing multiple sequence alignment (ClustalW).

**2.7 Mechanical Properties of Biofilm**

The co-cultured biofilms from the exposed Cu coupons (annealed Cu, 29.5% CW, and 56.2% CW) were removed from the corrosion cells at the end of the tests. The removed samples were stored under sterile conditions in a laminar air flow hood for 60 days. To analyze the intactness and durability of the biofilms, these samples were analyzed using MTS Nano Indenter XP equipped with a Berkovich indenter tip radius of 20 nm. These samples were mounted onto a stub with a low-temperature adhesive and introduced in a closed chamber to reduce noise levels. The test parameters included a maximum indenter depth of 2100 nm, a peak hold time of 5 s, a strain rate of 0.05 $s^{-1}$, and Poisson's ratio of 0.4 [27]. Fifteen indentations per sample were made, and the nanoindentation measurements were recorded in the form of a load-displacement curve. The hardness and the elastic modulus were determined using TestWorks4 software.

**2.8 Surface analysis and chemical composition**

The images of biofilm matrices on the exposed Cu samples, at the end of the corrosion tests, were acquired using a Zeiss Supra40 Scanning Electron Microscope (SEM) configured with Type II secondary electron (SE2). An accelerated voltage of 1 kV was used for the acquisition. The SEM was fitted with an Oxford Aztec Energy advanced system with Aztec 5.1 program to analyze energy dispersive spectroscopy (EDS) for elemental composition. The electron back-scattered diffraction was modified with an accelerated voltage of 15kV to achieve the EDS. The corrosion products were analyzed using an Ultima-Plus X-ray Diffractometer (XRD, Rigaku, Japan) with CuKα radiation configuration, scintillator counter, and a graphite monochromator. The Jade 7.5 software was to analyze XRD peaks. The thicknesses of the air-dried biofilm matrices were determined using VK-X250 laser confocal scanning microscope (CLSM) (Keyence Corp, Itasca, IL, USA). The CLSM images were obtained after Cu coupons were removed from the corrosion cells and stored under sterile conditions for 60 days. The average thickness was calculated using a multi-line roughness profile (4 lines in each sample), specifically by comparing the Cu substrate and the biofilm using VK-multifileanalyzer application software. A multi-line roughness profile was used for analyzing the pitting tendency of annealed Cu substrates exposed to the cultures of strain MIC21 and strain G20.

## 3. Results and Discussion

### 3.1 Ennoblement effect of the co-cultured biofilm

The three test reactors (annealed Cu, 29.5% CW, and 56.2% CW) initially experienced a decrease in the open circuit potential ($E_{ocp}$ vs. Ag/AgCl) (turned negative over time). This trend was restricted to the first 12-36 days (Figure 1), beyond which the $E_{ocp}$ values increased throughout the test duration. This ennoblement effect manifested passivating behavior of the biofilm matrix [28]. The $E_{ocp}$ values for annealed Cu decreased from -0.691 V to -0.799 V by day 12 and later took an opposite trend, and the values increased to -0.68 V by day 60. The $E_{ocp}$ values for 29.5% CW decreased from -0.733 V (day 0) to -0.746 V (day 24) and later increased to -0.599 V (day 60). The $E_{ocp}$ for 56.2% CW decreased from -0.741 V (day 0) to -0.788 V (day 36), followed by an increase to -0.635 V (day 60).

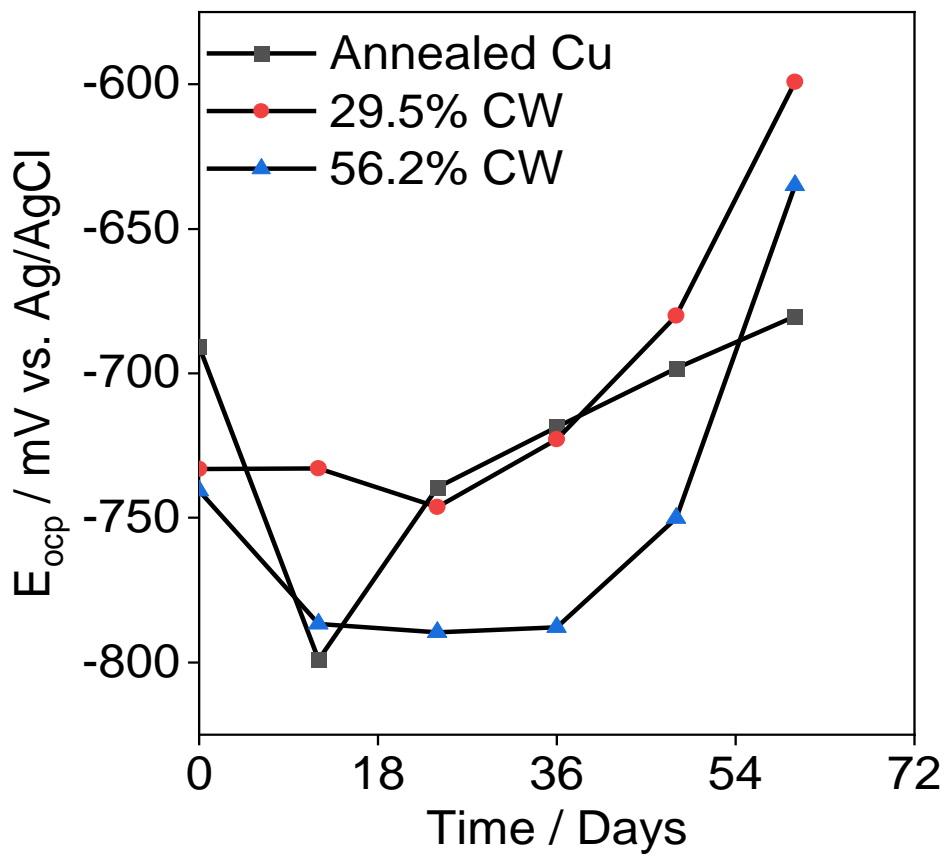

**Figure 1:** Open Circuit Potential profiles in the three corrosion cells.

The EIS results corroborated the ennoblement effect. The total impedance, represented by the diameter of the semicircle in the Nyquist plot, increased over time for annealed Cu (Figure 2a, shown for days 0, 30, and 60), 29.5% CW (Figure 2c), and 56.2% CW (Figure 2e), respectively. Higher impedance implies greater resistance to corrosion. Unlike Nyquist plots on day 0, the semicircles for Nyquist plots on days 30 and 60 did not extend fully into the low frequency. This observation further explains high impedance behavior during the latter times compared to day 0. The Bode plots also corroborated the passivation behavior in the three biocorrosion cells. The absolute value of impedance modulus at low frequency ($|Z|_{0.01Hz}$) on day 60 was 2.5 times higher than day 0 for annealed Cu (Figure 2b), 6.3 times higher for 29.2% CW (Figure 2d), and 4.2 times higher for 56.2% CW (Figure 2f). The resistance observed in the high-frequency region is attributed to the formation of a compact biofilm matrix (discussed in section 3.2.1) and in lower frequency resistance to relevant charge transfer reactions (details in Section 3.2).

The passivation behavior came as a surprise, considering our findings that the culture of *O. alaskensis* strain G20 aggravates the corrosion of underlying Cu surfaces. The additional test we performed for individual cultures of strain G20 corroborated the aggravation of corrosion on annealed Cu (Figure S1a) (section 3.3). In our earlier tests, we observed that the strain G20 cells promotes corrosion of Cu over time, irrespective of the protective coatings, which included polymer coatings and those modified with nanofillers [29] and nanoscale materials [2, 5, 30]. Prior studies also confirm that *Desulfovibrio* sp. promotes metallic corrosion under diverse environmental conditions [31]. A recently published review article by the authors' group provides comprehensive information on genes involved in biofilm formation and microbial corrosion that are shared across SRB genomes [32].

Limited studies have been done on the corrosion behavior of *Citrobacter* sp. *Citrobacter* sp. has been reported to promote MIC on X70 pipeline steel surfaces because of the metabolic activity of the bacteria in the early phase and the exfoliation and decomposition of the biofilm in the later phase of the exposure time [21]. However, the same study also elucidates the capability of *Citrobacter* sp. to form a compact biofilm matrix resulting in the ennoblement effect in OCP with an increase in charge transfer resistance and pore resistance during the midstage [21]. Zhang et al. also reported the ennoblement effect on OCP with microbiologically influenced corrosion inhibition when WC-Co was exposed to *Citrobacter* sp. in a sterilized emulsion [26]. Our study also showed the temporal increase in the impedance with the increase in the diameter of the Nyquist plot when the individual culture of *Citrobacter* sp. strain MIC21 was used under semi-aerobic conditions (Figure S1b) (section 3.3).

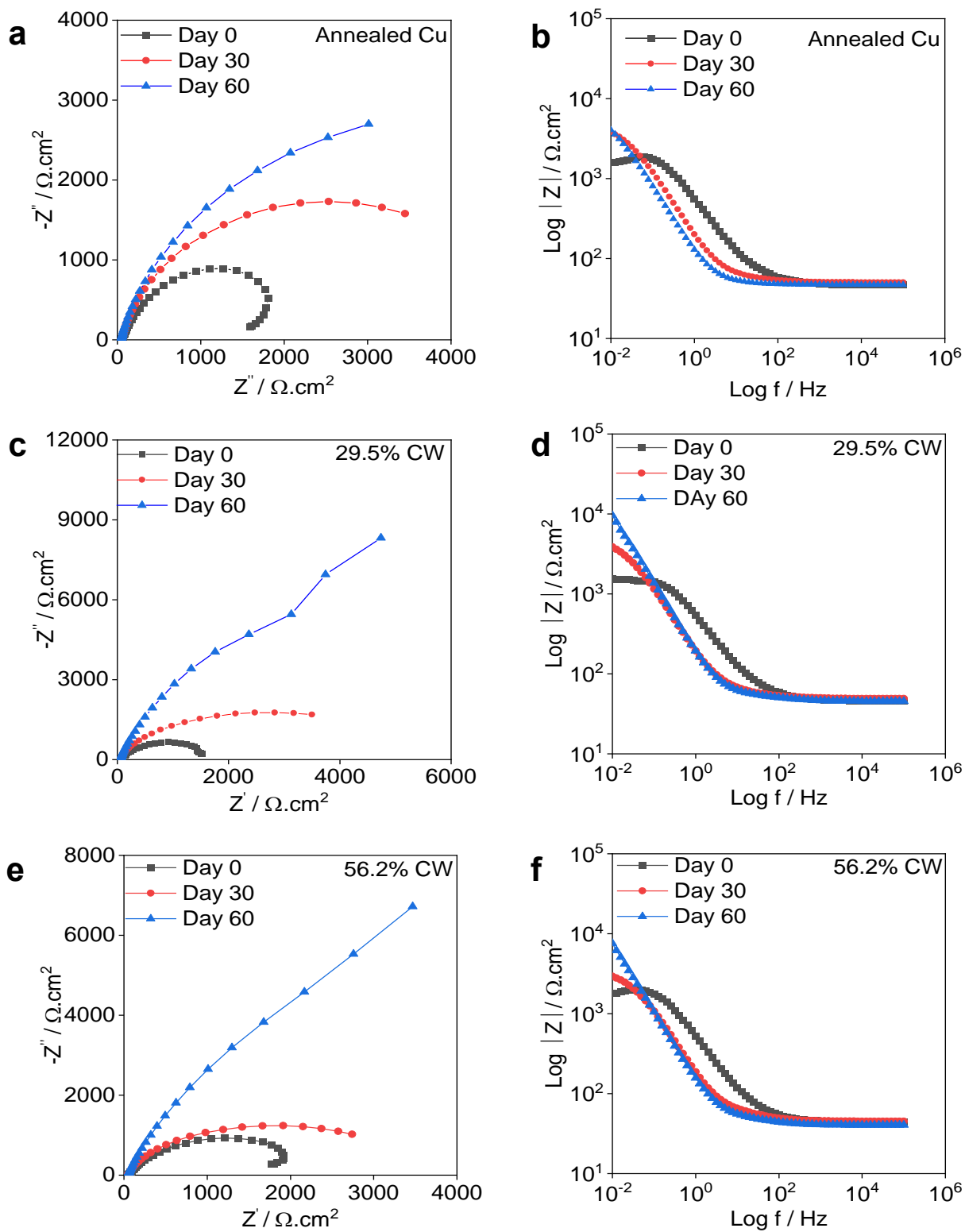

**Figure 2: Electrochemical Impedance Spectroscopy.** Nyquist and Bode plots for **(a)(b)** annealed copper, respectively, **(c)(d)** copper 29.5% CW, respectively, **(e)(f)** 56.2% CW copper, respectively.

## 3.2 Passivating mechanisms of *Citrobacter* sp. strain MIC21 in the co-cultured biofilm matrix

After establishing corrosion resistance of the biofilm matrix, we turn our attention to the underlying reasons for the passivation behavior. An electrical equivalent circuit (EEC) based on a modified Randles circuit (Figure 3a) was used to analyze EIS data from Nyquist and Bode plots (Figure 2). The goodness of fit was observed from the Chi-square parameters that ranged from $10^{-3}$ to $10^{-4}$, indicating a reasonable fit of the original data [33] (Table S4). The EEC analysis suggests that the passivation effects were primarily due to the barrier properties of biofilm matrices (Figure 3a).

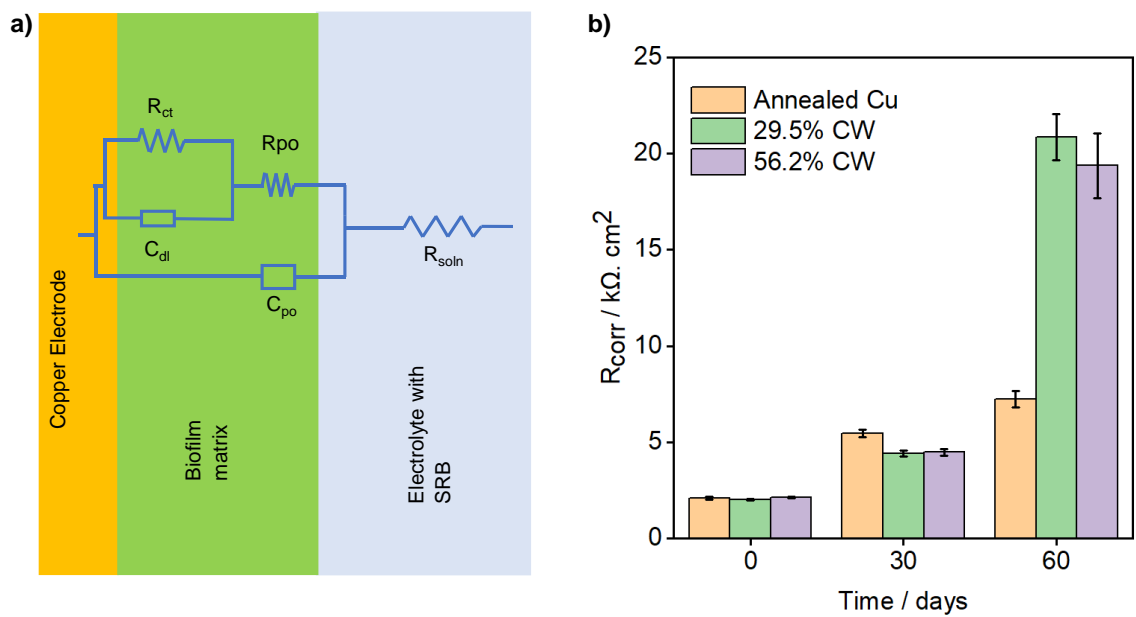

**Figure 3.** Electrical equivalent circuit (EEC) analysis of the biofilm matrix- (a) Electrical circuit with the corresponding physical model used for fitting the impedance spectra (b) Temporal profiles of total corrosion resistance for annealed Cu, 29.5% CW and 56.2% CW

The $R_{ct}$ values in the co-cultured test reactors decreased initially with a corresponding increase in corrosion rates. This anticipated behavior is due to the active metabolism of the proliferating cells [34] and inhomogeneities within the extracellular polymeric substance (EPS) film during the early stages of biofilm growth. The inhomogeneities in the EPS films allow corrosive ions (HS$^-$) to penetrate onto the Cu surfaces. Such penetration is evident from the formation of $Cu_2S$, a corrosion product (Eq 5) [35], as confirmed by the EDS and XRD analysis (See later sections).

A key finding here is regarding the passivation behavior, reflected in the form of increasing values of $R_{ct}$ and $R_{po}$ over time. The $R_{ct}$ for annealed Cu on day 60 (5.39 ±0.35 k$\Omega$.cm$^2$) was approximately 2.5-fold higher than day 0 (2.0±0.084 k$\Omega$.cm$^2$) (Table S4), and that for 29.5% CW on day 60 (9.84±0.29 k$\Omega$.cm$^2$) was approximately 5-fold greater than day 0 (1.86±0.025 k$\Omega$.cm$^2$) (Table S4). The $R_{ct}$ for 56.2% CW showed approximately 2-fold higher resistance from day 0 (2.02±0.023 kΩ.cm$^2$) to day 60 (6.57±0.48 kΩ.cm$^2$). We attribute this phenomenon to the formation of a compact biofilm matrix that served as a barrier to the penetration of any corrosive species [26]. The dense compact nature of the biofilm matrix is reflected by the temporal increase in the pore resistance for all three tests. For instance, the $R_{po}$ values increased by 18.5-fold on day 60 (1.85±0.086 kΩ.cm$^2$) compared to day 0 (100±7.86 Ω.cm$^2$) for annealed Cu. The $R_{po}$ for 29.5% and 56.2% CW increased by 65-fold and 113-fold, respectively, from day 0 to day 60 (Table S4). The significant increase in the pore resistance in CW samples can be attributed to the higher stress induced in the Cu coupons due to cold rolling and hence higher bacterial attachment. The higher the $R_{po}$, the greater is the resistance to penetration of HS$^-$.

The total corrosion resistance ($R_{corr} = R_{ct}+R_{po}$) for annealed Cu on day-60 (7.24±0.43 kΩ.cm$^2$) was nearly 3-fold higher than day 0 (2.1±0.092 kΩ.cm$^2$) (Figure 3b). Similarly, the resistances in the CW samples increased by at least 10-fold (Figure 3b). The higher resistance in the CW samples compared to annealed Cu can be attributed to the greater resiliency to overcome the higher degrees of stresses induced by the coldworking process.

**3.2.1 Domination of *Citrobacter* sp. strain MIC21 in the co-cultured biofilm matrices:** The gene sequencing studies revealed that *Citrobacter* sp. strain MIC21 outcompeted *O.alaskensis* strain G20 at the end of the co-cultured corrosion tests. The biofilm samples at the end of the tests showed 98.93% identity (with 99% query coverage) with *Citrobacter freundii* FC18565 (Acc No: MK561018.1), based on molecular identification by 16S rRNA gene sequencing and BLAST analysis (Figure S2). These samples were quantified based on their 16S rRNA gene copies in log10 gene copies/μL. Phylogenetic analysis of 16S rRNA gene sequences of strain MIC21 showed the evolutionary relationship between similar species. The neighbor-joining method showed the evolutionary relationship of MIC21 with other *Citrobacter* sp. The 16S rRNA gene sequences were submitted to GenBank (accession number: OK144236). The SEM images revealed rod-shaped bacterium (L = 2-3 µm; W = ~500 nm) with a smoother cell wall (Figure 5a, b, c), reflecting morphological features of Gram-negative bacterial species.

Our gas chromatography tests revealed the presence of minimal levels of oxygen (~14%) in the headspace. The presence of this oxygen can provide a selective advantage for the early colonization by *Citrobacter* sp. strain MIC21 over *O. alaskensis* strain G20. This situation allows MIC21 cells to use oxygen as an electron acceptor [36], allowing them to prolong their generation time and lag phase [37]. Upon depletion of the oxygen, strain MIC21 cells shift into a dissimilatory sulfate reduction pathway, which is the only possible pathway for *O. alaskensis* strain G20, which is an obligate anaerobe. Literature has shown *Citrobacter* sp.'s ability to grow faster in an aerobic condition and quickly recover to reduce sulfate when transferred from an aerobic to an anaerobic environment [38]. The dissimilatory reduction pathways generate corrosive metabolites such as $HS^-$ [39]. However, *Citrobacter* sp. tolerates higher levels of toxic Cu species compared to *Desulfovibrio vulgaris* [39].

**3.2.2 Compact nature of the *Citrobacter* sp. strain MIC21 biofilm matrices under the co-cultured condition:** The microscopy studies at the end of the corrosion tests revealed a compact biofilm comprising homogeneous layers of the rod-shaped *Citrobacter* sp. strain MIC21 cells. This compact nature was consistently observed on annealed Cu, 29.5% CW, and 56.2% CW (Figure 4a, 4b, 4c, respectively). Their average biofilm thicknesses were 29.07±1.45 µm, 43.40±2.88 µm, and 26.24±1.79 µm, respectively (Figure S3a-S3c), that are significantly greater than a typical thickness of *Oleidesulfovibrio alaskensis* strain G20 (0-15 µm) [5] and other *Desulfovibrio sp.* (~20 µm) [40]. The morphological features of the biofilms observed in the current study were significantly different compared to *O. alaskensis* strain G20 based corrosion systems in our current and earlier studies [2, 30]. The mean hardness of the biofilms for annealed Cu, 29.5% CW, and 56.2% CW (Figures 4d-4e) were as high as 12.7±2.4 MPa, 5.2±1.8 MPa, and 4.5±2.1 MPa, respectively. Their elastic modulus values ranged from 0.4-8 GPa (Figure 4e), which are ~4000-fold higher than typical SRB biofilms [41]. A higher modulus indicates higher stiffness of the biofilm matrix composition with less elasticity. The indentation depth was determined to be below 10% of the average biofilm thickness for all three samples. The load/displacement curve (Figure 4d) shows the change in the curve pattern for all three samples at different displacements. We didn't observe any biofilm formation on the Cu samples before the samples were exposed to co-cultured media (Figures S4a, S4c, S4e).

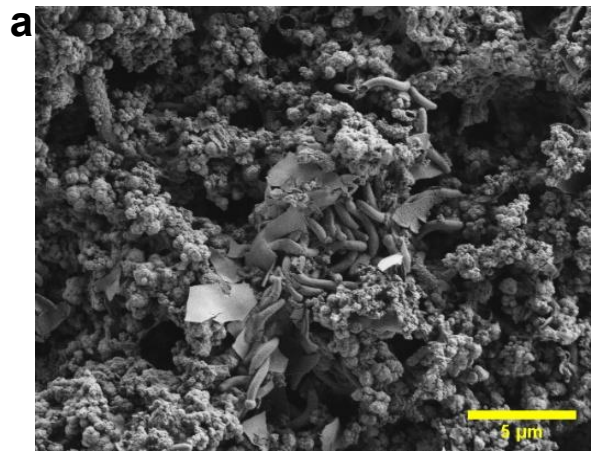
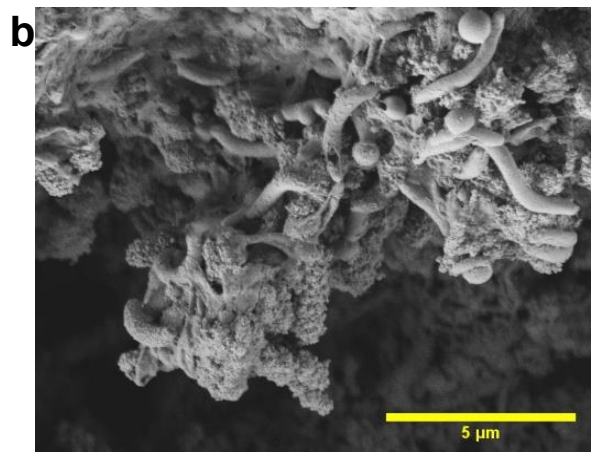
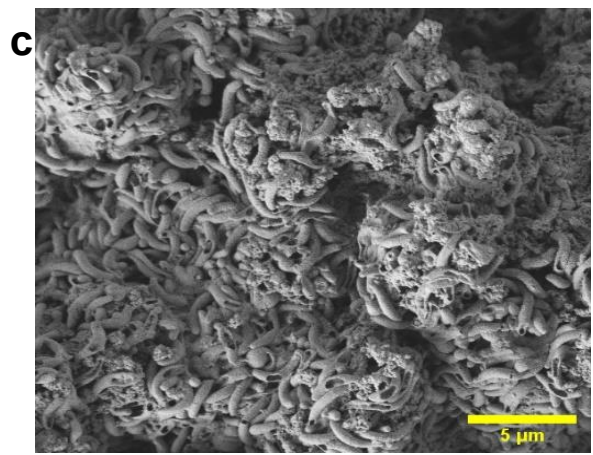
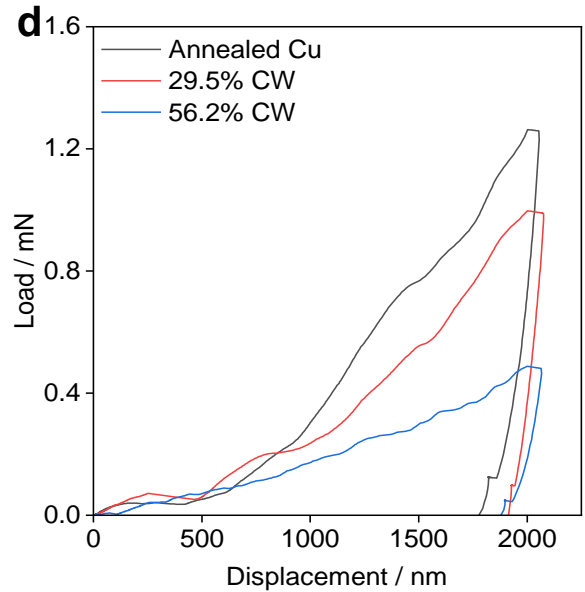
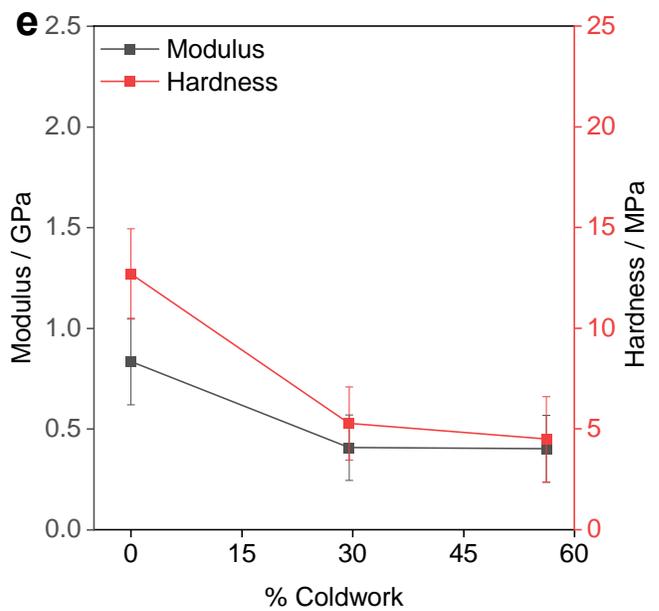

**Figure 4: Compact biofilm matrices and their mechanical properties.** SEM images of **(a)** annealed Cu, **(b)** 29.5% CW, and **(c)** 56.2% CW after 60 days of exposure. Nanoindentation tests **(d)** Load vs. displacement curves of biofilms on different percent CW samples, **(e)** Biofilm hardness/modulus with displacement at maximum load as a function of percent CW samples.

We analyzed the corrosion products in the biofilm matrices using the EDS and XRD tests. The EDS analysis revealed sharp Cu peaks (Figure S5) for pristine Cu samples; specifically, they displayed Cu (111), (200), (220), (311), and (222) orientations (Figures S4b, S4d, S4f). However, the exposed annealed samples developed signatures of carbon (C), sulfur (S), and oxygen (O) (Figure 5a) with compositions of 14.7%, 13.7%, and 11.5% w/w, respectively (Figure S6a). The C signatures are attributed to the EPS products, and S and O signatures to the corrosion products. The 29.5% CW (Figure 5c) showed 21.4% O, 13.2% C, and 11.4% S, respectively (Figure S6b). The 56.2% CW sample (Figure 5e) displayed 15.4% O, 13.5% S, and 12.2% C, respectively (Figure S6c).

The XRD analyses confirmed two key signatures of MIC, namely chalcocite ($Cu_2S$) and copper oxide (CuO), in all three corrosion tests, namely annealed Cu (Figure 5b), 29.5% CW (Figure 5d), and 56.2% CW (Figure 5f). The EPS components captured a significant amount of Cu (annealed Cu:57.4% ; 29.5% CW: 44.4% ; 56.2% CW: 56.4% ), thus limiting the mass transfer within the biofilm matrix [42]. The CuO and $Cu_2S$ signatures observed in the biofilm matrices (Figure 5) represent active constituents of passivating layers [43] that resist the diffusion of corrosive ions [44].

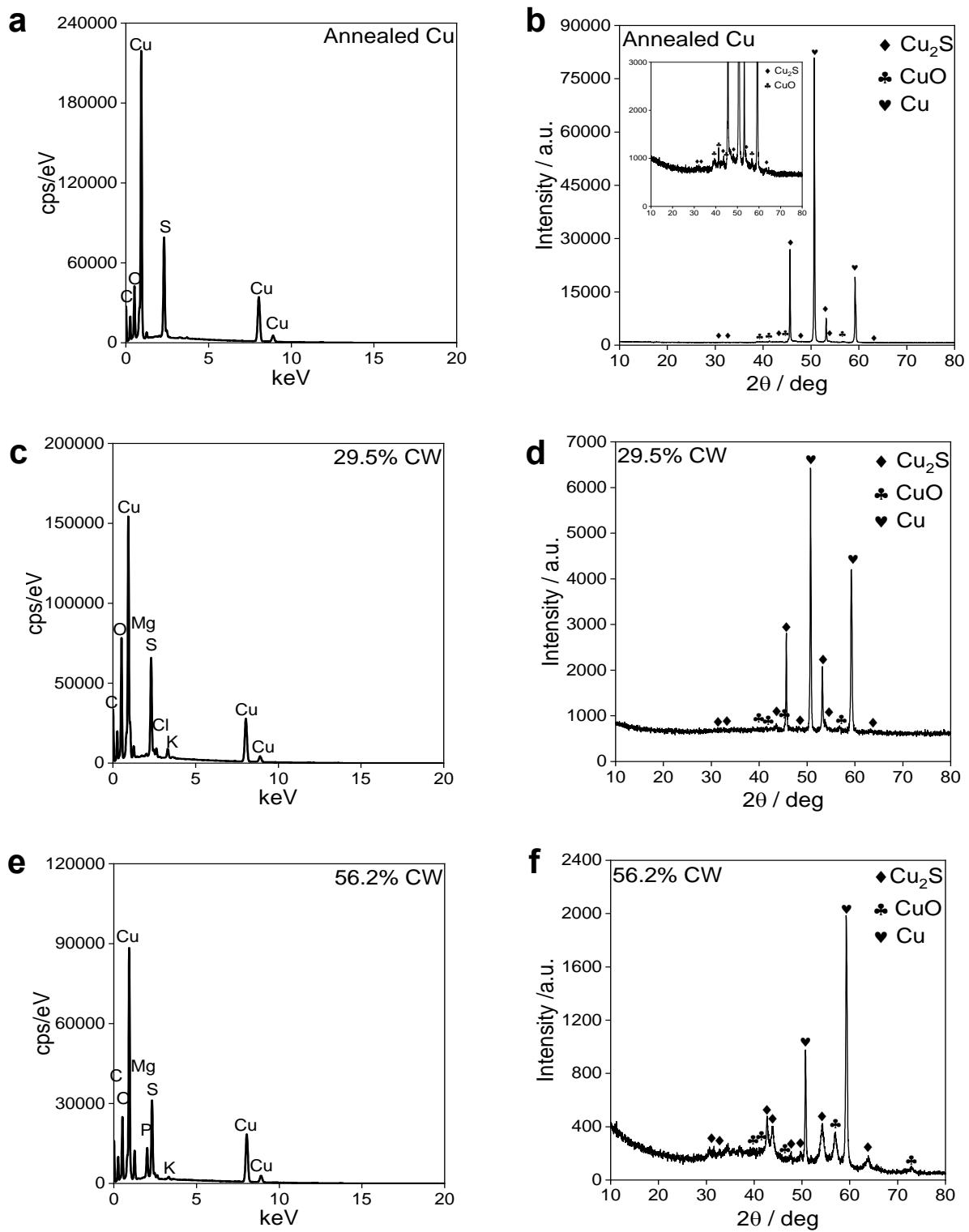

**Figure 5: Composition of biofilm matrix**. EDS data showing the elemental composition for **(a)** annealed Cu, **(c)** 29.5% CW, and **(e)** 56.2% CW; XRD data showing the composition of corrosion products for **(b)** annealed Cu, **(d)** 29.5% CW, and **(f)** 56.2% CW, respectively showing $Cu_2S$ and CuO.

### 3.3. Additional tests with individual cultures:

The doubling time of strain MIC21 under the aerobic condition was found to be 13.10 hrs, the anaerobic condition to be 14.9 hrs, and strain G20 under the anaerobic condition to be 17.5 hrs (Figure S7). This explains the capability of strain MIC21 to colonize and form biofilm faster than strain G20.

The passivation effect was further corroborated by observing the electrochemical behavior of individual cultures of strain MIC21 compared with strain G20. The diameter of the semi-circle in the Nyquist plot had a temporal increment for strain MIC21 (Figure S1a), whereas the strain G20 showed the opposite trend (Figure S1b). The total corrosion resistance ($R_{corr} = Rct + Rpo$) increased by approximately 4-fold from day 0 (15.62±0.49 kΩ. cm$^2$) to day 15 (59.0±11.04 kΩ. cm$^2$) (Table S5). As expected, *O. alaskensis* strain G20 showed a reduction in the $R_{corr}$ by nearly 6-fold from 5.50±1.3 kΩ. cm$^2$ on day 0 to 0.9±0.022 kΩ. cm$^2$ on day 15 (Table S5).

The direct evidence for corrosion resistance by *Citrobacter* sp. strain MIC21 was achieved from Tafel analysis after 15 days of corrosion test. Strain MIC21 offered corrosion resistance with a rate of 0.5 mpy, approximately 2.5-fold lesser than the resistance offered by strain G20 (1.18 mpy) on annealed copper (Figure S1c). The corrosion rate after 60 days of the test for co-cultured condition showed 0.399 mpy (Figure S1d) for annealed Cu. The weight loss measurement substantiated the results from the Tafel analysis (Figure S8). The corrosion rate of annealed Cu exposed to co-culture (0.430±0.068 mpy) was approximately 1-fold less than the individual culture of strain MIC21 (0.445±0.005 mpy) and approximately 3-fold less than the individual culture of strain G20 (1.371±0.098 mpy).

The pitting profiles also demonstrated the corrosion resistance behavior of strain MIC21 when compared with strain G20. Annealed Cu exposed to the strain G20 showed several pits (Figure S9) with an average peak depth of 0.41±0.10 µm and a maximum depth of 0.54 µm, implying severe corrosion. However, the peak density on Cu coupons exposed to strain MIC21 was significantly less, with an average peak depth of 0.25±0.07 µm and a maximum depth of 0.32 µm.

**3.4 Outlook**: Members of gamma-proteobacteria have been reported to outcompete their peers in response to stressful environmental conditions resistance mechanisms. For instance, given a co-culture of *Pseudomonas aeruginosa* and *Vibrio cholerae* [45], *P. aeruginosa* has been reported to induce an antibacterial attack on *V. cholerae* in response to the toxic compounds secreted by type 6 secretion system (T6SS) of *V. cholerae*. The latter process is referred as T6SS dueling effect. Along these lines, *Citrobacter* sp. strain MIC21 outcompeted *Oleidesulfovibrio alaskensis* strain G20 in response to the presence of toxic copper species from the dissimilatory sulfate reduction pathways of *O.alaskensis* strain G20 [37]. The *Citrobacter* sp. strains harbor plasmid-mediated quinolone resistance genes and beta-lactamase gene sets that display virulent mechanisms [46]. Exposure to heavy metals has also been reported to proliferate antibiotic resistance genes (ARGs). *Omics* studies are warranted to understand the mechanisms of metal resistance, antimicrobial resistance, and ARG that allowed the out competition of *Citrobacter* sp. Such Omics studies can also reveal the roles of co-resistance and cross-resistance mechanisms in response to selection pressure induced by the metal exposure [47].

## 4. Conclusion

This study demonstrates the use of the biofilm matrix of *Citrobacter* sp. as a protective coating for microbial corrosion prevention. The protection ability was demonstrated for annealed copper as well as those modified with stresses. The MIC resistance of the biofilm matrix was manifested in the form of a compact and thick barrier film that prohibitively restricted the penetration of corrosion ions. For instance, such compact biofilms are known to yield cathodic inhibitors based on biopolymers from EPS that convert Cu ions into insoluble precipitates and block active sites of the underlying copper surfaces. The evolution of such a compact biofilm was attributed to the bacterial stress response that allowed *Citrobacter* sp. (facultative anaerobe) to outcompete its counterpart (obligate anaerobe *O.alaskensis*). Omics studies are warranted to understand the specific stress resistance mechanisms of *Citrobacter* sp. at a genetic level. The findings revealed in the current study can be used to design minimally invasive approaches for invoking *Citrobacter* sp. metabolism and, in turn, alleviate undesirable effects of sulfate reducing bacteria, especially obligate anaerobes, involved in the corrosion of metals. Further studies are warranted to reliably transfer large area biofilm matrices of *Citrobacter* sp. onto arbitrary metal substrates and test their long-term protection ability.

## 5. CRediT authorship contribution statement

**Pawan Sigdel**: Resources, Conceptualization, Visualization, and Validation. **Ananth Kandadai: Validation. Kalimuthu Jawaharraj:** Validation. **Bharat Jasthi :** Resources, Funding acquisition. **Venkataramana Gadhamshetty**: Resources, Conceptualization, Supervision, Validation, Project administration, Funding acquisition.

## 6. Declaration of Competing Interest

The authors declare that they have no known competing financial interests or personal relationships that could have appeared to influence the work reported in this paper.

## 7. Acknowledgments

We acknowledge the funding support from National Science Foundation RII FEC awards (#1849206, #1920954) and support from the Department of Civil and Environmental Engineering at the South Dakota Mines.

## 8. Appendix A. Supplementary data

Supplementary data to this article can be found online

## 9. References

1. Alasvand Zarasvand, K. and V. Ravishankar Rai, *Identification of the traditional and non-traditional sulfate-reducing bacteria associated with corroded ship hull.* 3 Biotech, 2016. **6**(2): p. 197-197.
2. Chilkoor, G., et al., *Atomic Layers of Graphene for Microbial Corrosion Prevention.* ACS Nano, 2021. **15**(1): p. 447-454.
3. Chen, S., P. Wang, and D. Zhang, *Corrosion behavior of copper under biofilm of sulfate-reducing bacteria.* Corrosion Science, 2014. **87**: p. 407-415.
4. Licina, G.J. and D. Cubicciotti, *Microbial-induced corrosion in nuclear power plant materials.* JOM, 1989. **41**(12): p. 23-27.
5. Chilkoor, G., et al., *Hexagonal boron nitride for sulfur corrosion inhibition.* ACS nano, 2020. **14**(11): p. 14809-14819.
6. Huttunen-Saarivirta, E., P. Rajala, and L. Carpén, *Corrosion behaviour of copper under biotic and abiotic conditions in anoxic ground water: electrochemical study.* Electrochimica Acta, 2016. **203**: p. 350-365.
7. Kakooei, S., M.C. Ismail, and B. Ariwahjoedi, *Mechanisms of microbiologically influenced corrosion: a review.* World Appl. Sci. J, 2012. **17**(4): p. 524.
8. Gu, T., et al., *Toward a better understanding of microbiologically influenced corrosion caused by sulfate reducing bacteria.* Journal of materials science & technology, 2019. **35**(4): p. 631-636.
9. Li, Y., et al., *Anaerobic microbiologically influenced corrosion mechanisms interpreted using bioenergetics and bioelectrochemistry: a review.* Journal of Materials Science & Technology, 2018. **34**(10): p. 1713-1718.
10. Bard, A.J. and L.R. Faulkner, *Electrochemical Methods: Fundamentals and Applications.* Surface Technology, 1983. **20**(1): p. 91-92.

11. Grass, G., C. Rensing, and M. Solioz, *Metallic copper as an antimicrobial surface.* Applied and environmental microbiology, 2011. **77**(5): p. 1541-1547.
12. Chen, S. and D. Zhang, *Study of corrosion behavior of copper in 3.5 wt.% NaCl solution containing extracellular polymeric substances of an aerotolerant sulphate-reducing bacteria.* Corrosion Science, 2018. **136**: p. 275-284.
13. Qi, H., et al., *Bioinspired multifunctional protein coating for antifogging, self-cleaning, and antimicrobial properties.* ACS applied materials & interfaces, 2019. **11**(27): p. 24504-24511.
14. Makhlouf, A.S.H. and N.Y. Abu-Thabit, *Advances in smart coatings and thin films for future Industrial and Biomedical Engineering Applications*. 2019: Elsevier.
15. Lou, Y., et al., *Microbiologically influenced corrosion inhibition mechanisms in corrosion protection: A review.* Bioelectrochemistry, 2021. **141**: p. 107883.
16. Jayaraman, A., et al., *Axenic aerobic biofilms inhibit corrosion of SAE 1018 steel through oxygen depletion.* Applied microbiology and biotechnology, 1997. **48**(1): p. 11-17.
17. Dubiel, M., et al., *Microbial iron respiration can protect steel from corrosion.* Applied and environmental microbiology, 2002. **68**(3): p. 1440-1445.
18. Bai, L., et al., *Isolation and characterization of cytotoxic, aggregative Citrobacter freundii.* PLoS One, 2012. **7**(3): p. e33054.
19. Alasvand Zarasvand, K. and V. Ravishankar Rai, *Identification of the traditional and non-traditional sulfate-reducing bacteria associated with corroded ship hull.* 3 Biotech, 2016. **6**(2): p. 1-8.
20. Yan, J., et al., *Carbon metabolism and sulfate respiration by a non-conventional Citrobacter freundii strain SR10 with potential application in removal of metals and metalloids.* International Biodeterioration & Biodegradation, 2018. **133**: p. 238-246.
21. Shahryari, Z., K. Gheisari, and H. Motamedi, *Effect of sulfate reducing Citrobacter sp. strain on the corrosion behavior of API X70 microalloyed pipeline steel.* Materials Chemistry and Physics, 2019. **236**: p. 121799.
22. Zhao, C., et al., *Isolation of a sulfate reducing bacterium and its application in sulfate removal from tannery wastewater.* African Journal of Biotechnology, 2011. **10**(56): p. 11966-11971.
23. Hu, J., et al., *Increased excess intracellular cyclic di-AMP levels impair growth and virulence of Bacillus anthracis.* Journal of bacteriology, 2020. **202**(9): p. e00653-19.
24. NACE, A., *Standard guide for laboratory immersion corrosion testing of metals.* ASTM Int., 2012: p. 1-9.
25. G1, A. *Standard Practice for Preparing, Cleaning, and Evaluating Corrosion Test Specimens*. ASTM West Conshohocken, PA.
26. Zhang, Q., et al., *Corrosion behavior of WC–Co hardmetals in the oil-in-water emulsions containing sulfate reducing Citrobacter sp.* corrosion science, 2015. **94**: p. 48-60.
27. Kundukad, B., et al., *Mechanical properties of the superficial biofilm layer determine the architecture of biofilms.* Soft matter, 2016. **12**(26): p. 5718-5726.
28. George, R., et al., *Microbiologically influenced corrosion of AISI type 304 stainless steels under fresh water biofilms.* Materials and Corrosion, 2000. **51**(4): p. 213-218.
29. Chilkoor, G., et al., *Maleic anhydride-functionalized graphene nanofillers render epoxy coatings highly resistant to corrosion and microbial attack.* Carbon, 2020. **159**: p. 586-597.
30. Chilkoor, G., et al., *Hexagonal Boron Nitride: The Thinnest Insulating Barrier to Microbial Corrosion.* ACS Nano, 2018. **12**(3): p. 2242-2252.
31. Ilhan-Sungur, E., N. Cansever, and A. Cotuk, *Microbial corrosion of galvanized steel by a freshwater strain of sulphate reducing bacteria (Desulfovibrio sp.).* Corrosion Science, 2007. **49**(3): p. 1097-1109.
32. Kumar Tripathi, A., et al., *Gene sets and mechanisms of sulfate-reducing bacteria biofilm formation and quorum sensing with impact on corrosion.* Frontiers in microbiology, 2021: p. 3120.

33. Unsal, T., et al., *Effects of Ag and Cu ions on the microbial corrosion of 316L stainless steel in the presence of Desulfovibrio sp.* Bioelectrochemistry, 2016. **110**: p. 91-99.
34. Guan, F., et al., *Influence of sulfate-reducing bacteria on the corrosion behavior of 5052 aluminum alloy.* Surface and Coatings Technology, 2017. **316**: p. 171-179.
35. Yuan, S., et al., *Surface chemistry and corrosion behaviour of 304 stainless steel in simulated seawater containing inorganic sulphide and sulphate-reducing bacteria.* Corrosion Science, 2013. **74**: p. 353-366.
36. Shanks, R.M., et al., *Isolation and identification of a bacteriocin with antibacterial and antibiofilm activity from Citrobacter freundii.* Archives of microbiology, 2012. **194**(7): p. 575-587.
37. Wang, X., et al., *Coupling heavy metal resistance and oxygen flexibility for bioremoval of copper ions by newly isolated Citrobacter freundii JPG1.* Journal of environmental management, 2018. **226**: p. 194-200.
38. Liu, Z.-h., et al., *Sulfate-reducing bacteria in anaerobic bioprocesses: basic properties of pure isolates, molecular quantification, and controlling strategies.* Environmental Technology Reviews, 2018. **7**(1): p. 46-72.
39. Qiu, R., et al., *Sulfate reduction and copper precipitation by a Citrobacter sp. isolated from a mining area.* Journal of Hazardous Materials, 2009. **164**(2-3): p. 1310-1315.
40. Chen, L., B. Wei, and X. Xu, *Effect of sulfate-reducing bacteria (SRB) on the corrosion of buried pipe steel in acidic soil solution.* Coatings, 2021. **11**(6): p. 625.
41. Huang, C., et al., *Effect of Nonphosphorus Corrosion Inhibitors on Biofilm Pore Structure and Mechanical Properties.* Environmental science & technology, 2020. **54**(22): p. 14716-14724.
42. Vargas, I.T., et al., *Copper corrosion and biocorrosion events in premise plumbing.* Materials, 2017. **10**(9): p. 1036.
43. Videla, H.A. and L.K. Herrera, *Understanding microbial inhibition of corrosion. A comprehensive overview.* International Biodeterioration & Biodegradation, 2009. **63**(7): p. 896-900.
44. Hernandez, G., et al., *Corrosion inhibition of steel by bacteria.* Corrosion, 1994. **50**(8): p. 603-608.
45. Basler, M., B. Ho, and J. Mekalanos, *Tit-for-tat: type VI secretion system counterattack during bacterial cell-cell interactions.* Cell, 2013. **152**(4): p. 884-894.
46. Liu, L., et al., *Antimicrobial resistance and cytotoxicity of Citrobacter spp. in Maanshan Anhui Province, China.* Frontiers in microbiology, 2017. **8**: p. 1357.
47. Baker-Austin, C., et al., *Co-selection of antibiotic and metal resistance.* Trends in microbiology, 2006. **14**(4): p. 176-182.

# Supplementary Information
# Microbial Corrosion Prevention by *Citrobacter* sp. Biofilms

Pawan Sigdel[1,3], Ananth Kandadai[2,3], Kalimuthu Jawaharraj[1,3,4], Bharat Jasthi[2,3,4], Etienne Gnimpieba[3,4,5], Venkataramana Gadhamshetty[1,3,4*]

[1]Civil and Environmental Engineering, South Dakota School of Mines and Technology, 501 E. St. Joseph Street, Rapid City, SD, 57701, USA

[2]Materials and Metallurgical Engineering, South Dakota School of Mines and Technology, 501 E. St. Joseph Street, Rapid City, SD, 57701, USA

[3]2D-materials for Biofilm Engineering, Science and Technology (2DBEST) Center, South Dakota School of Mines and Technology, 501 E. St. Joseph Street, Rapid City, SD, 57701, USA

[4]Data-Driven Materials Discovery for Bioengineering Innovation Center, South Dakota Mines, 501 E. St. Joseph Street, Rapid City, SD, 57701, USA

[5]Biomedical Engineering, University of South Dakota, 4800 N Career Ave, Sioux Falls, SD 57107, USA

[*]**Corresponding author.**

*E-mail address:* Venkata.Gadhamshetty@sdsmt.edu (V. Gadhamshetty).

Table S1. An overview of MIC studies on copper by SRB, NP: Not provided

| # Cu substrate (SRB name) | MIC mechanism | References |
|---|---|---|
| Cu(110) (Cu > 99.9%, =0.04% O, remaining trace elements, mass%) (*Desulfovibrio Vulgaris*) | MIC caused by sulfide secreted by SRB | [1] |
| Cu (> 99.9% mass%; diameter =10 mm, and thickness = 4 mm)(NP) | Biofilm was formed by the SRB, and cuprous sulfide was the major corrosion product. Corrosion rate directly affected by SRB growth cycle and metabolism with localized corrosion during exponential and stationary phase. | [2] |
| 70Cu-30Ni alloy(NP) | Intergranular corrosion after 7 days of immersion in sea water containing SRB. Metabolites concentration coincided with bacterial growth, metal sulfides produced from copper and nickel | [3] |
| Cu cylinders (>99.9%, mass%; diameter = 0.5 cm diameter, height = 0.5 cm) (*Desulfovibrio* Sp.) | Extracellular Polymeric Substance of SRB exposed to Cu on 3.5 wt% NaCl showed Cu corrosion inhibition for a short term. Longer immersion time promoted corrosion by degrading protective $Cu_2O$ film | [4] |

**Table S2.** An overview of Microbial Induced Corrosion Inhibition (MICI) mechanisms

| Bacteria *(strain)* | Characteristics | Function | References |
|---|---|---|---|
| *Serratia marcescens* EF 190, *Pseudomonas* sp. *S9,* | Aerobic | Aerobic respiration resulting in lo-oxygen barrier inhibiting corrosion of steel. | [5] |
| *Staphylococcus* sp. | Facultative anaerobe | EPS composed of hydrophobic components formed a corrosion inhibition barrier on a low carbon steel surface. | [6] |
| *B. subtilis* | Facultative anaerobe | Dense, uniform, and hydrophobic biofilm composed of polysaccharides/TasA amyloid fibers in low-alloy steel. | [7] |
| SRB LVform6 | Anaerobic | Dolomite precipitation under low temperatures and anoxic conditions results in corrosion inhibition | [8] |
| Nitrate Reducing Bacteria | Facultative anaerobe | In the presence of nitrate, NRB inhibits SRB growth and thus reduces $H_2S$ production | [9] |
| SRB *Citrobacter freundii* | Facultative anaerobe | Citrobacter spp. Produced stronger biofilm compared to *Desulfovibrio* spp., lowest mass loss was observed in *C.freundii*. | [10] |

**Table S3**. Corrosion behavior of *Citrobacter* Species

| # | Strain | Metal Substate | Major Findings | References |
|---|--------|----------------|----------------|------------|
| 1 | *Citrobacter farmeri* | Q235 Carbon steel coupons | OCV values shifted negative and continued to become more negative than artificial seawater until 48h implying discontinuous biofilm formation and microbial colonization with acidic metabolite production. After 72 h, a noble shift attributed to protective and compact biofilm formation. | [11] |
| 2 | *Citrobacter* Sp. | API X70 microalloyed pipeline steel | Significant reduction in charge transfer resistance from 400Ωcm$^2$ after day 7 to 55 Ωcm$^2$ after day 21. Metabolism of bacteria and formation of heterogeneously dispersed compact biofilm was attributed for reduced resistance. | [12] |
| 3 | *Citrobacter koseri* | Tungsten carbide cobalt in oil-in-water emulsion | Microbial influence corrosion of WC-30C$_0$ in O/W emulsion and nutrient was attributed to Citrobacter sp. Citrobacter sp. containing emusion showed microbiologically influenced corrosion inhibition. | [13] |

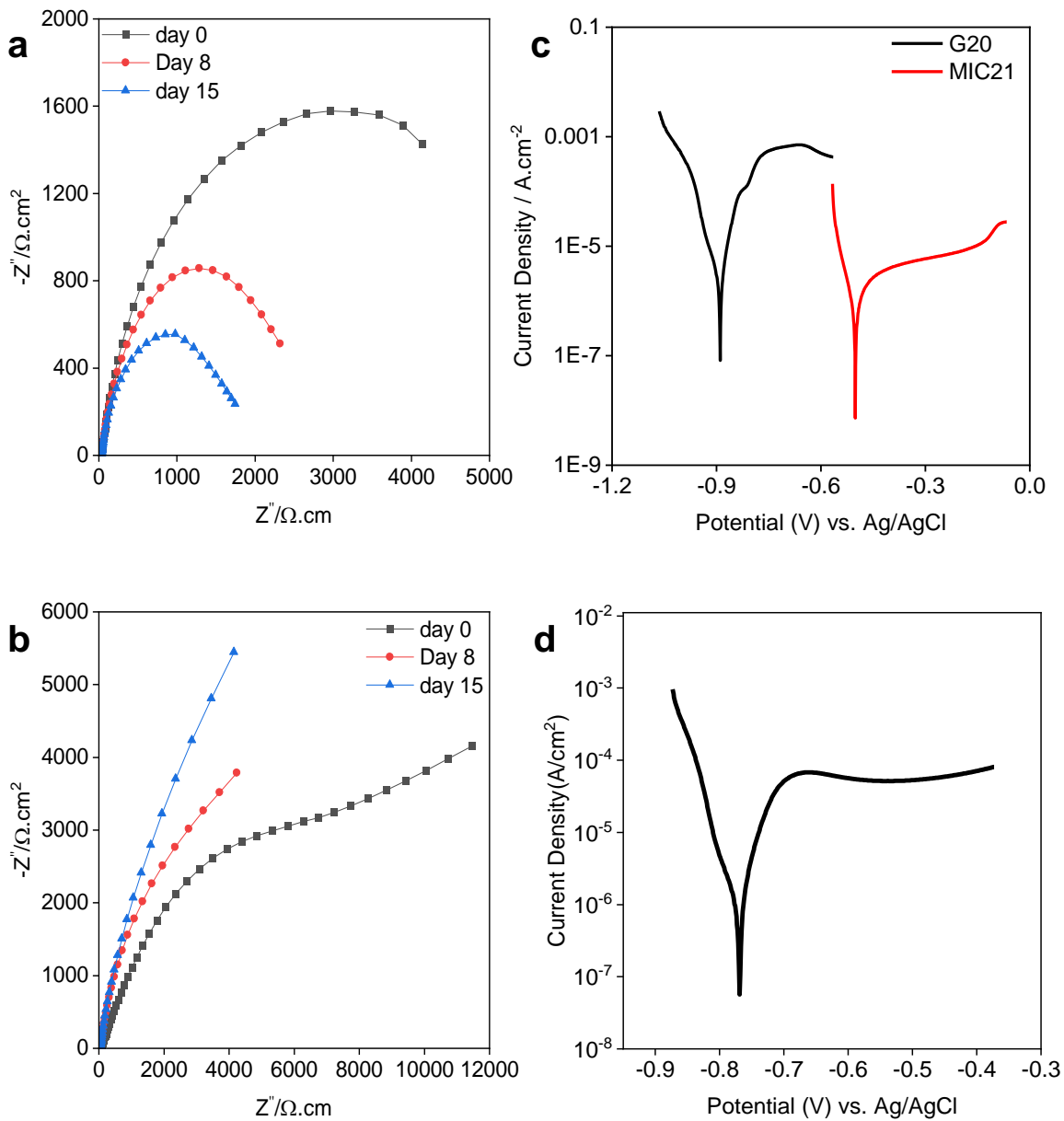

**Figure S1 Electrochemical Impedance Spectroscopy (EIS) and Tafel analysis** - (a) Nyquist plot for copper annealed Cu showing the increase in impedance from day 0 to 15 when exposed to individual culture of strain G20 (b) Nyquist plot for annealed Cu showing the decrease in impedance from day 0 to day 15 when exposed to individual culture of strain MIC21, (c) Tafel analysis shown higher corrosion rate for annealed Cu when exposed to strain G20 in comparison to strain MIC21, (d) Tafel analysis of annealed Cu when exposed to co-culture of strain G20 and MIC21 for 60 days.

**Table S4.** Fitting results of EIS for co-cultured media in annealed Cu, 29.5 % CW and 56.2% CW during 60 days of corrosion test

| Days | $R_{sol}$ ($\Omega.cm^2$) | $R_{ct}$ ($\Omega.cm^2$) | $R_{po}$ ($\Omega.cm^2$) | $C_{dl}$ (F cm$^{-2}$) | $C_{po}$ (F cm$^{-2}$) | $R_{corr}$ ($\Omega.cm^2$) | Goodness of Fit |
|---|---|---|---|---|---|---|---|
| Annealed Cu | | | | | | | |
| 0 | 47±0.32 | 2000±84.44 | 100±7.861 | 0.000166±0.001164 | 0.0002392±0.00011 | 2100±92.30 | 0.005978 |
| 30 | 51.62±0.29 | 4473±122.3 | 1000±63.94 | 0.00005836±0.0000219 | 0.001141±0.000018 | 5473±186.24 | 0.0009278 |
| 60 | 47.13±0.25 | 5390±350.43 | 1850±86.43 | 0.00000983±0.0000012 | 0.001847±0.000014 | 7240±436.86 | 0.0004023 |
| 29.5% Cu | | | | | | | |
| 0 | 46.61±0.30 | 1860±25.74 | 170±16.80 | 0.000128±0.0000149 | 0.0004156±0.000010 | 2030±42.54 | 0.001988 |
| 30 | 50.41±0.28 | 3570±85.79 | 850±78.16 | 0.0000902±0.000016 | 0.00111±0.0000094 | 4420±163.95 | 0.001452 |
| 60 | 47.5±0.26 | 9840±297.56 | 11000±905 | 0.00002765±0.0000071 | 0.001089±0.000018 | 20840±1202.56 | 0.002786 |
| 56.2% CW | | | | | | | |
| 0 | 44.67±0.30 | 2020±23.18 | 113±22 | 0.00002±0.0000029 | 0.0004026±0.0000055 | 2130±45.18 | 0.003544 |
| 30 | 46.49±0.25 | 3274±111.4 | 1220±62.81 | 0.00011±0.000012 | 0.001262±0.000011 | 4496±174.21 | 0.002556 |
| 60 | 42.08±0.24 | 6570±486 | 12800±1155 | 0.00003954±0.0000012 | 0.001404±0.000010 | 19370±1641 | 0.003298 |

**Table S5.** Fitting results of EIS for individual cultures in annealed Cu during 15 days of corrosion test

| Days | $R_{ct}$ (k$\Omega$.cm$^2$) | $R_{po}$ (k$\Omega$.cm$^2$) | $C_{dl}$ (F.cm$^{-2}$) | $C_{po}$ (F.cm$^{-2}$) | Total corrosion resistance ($R_{ct}+R_{po}$) (k$\Omega$.cm$^2$) | Goodness of fit |
|---|---|---|---|---|---|---|
| **MIC21/Annealed Cu** | | | | | | |
| 0 | 11.0±0.25 | 4.62±0.24 | 1.23E-06±561.8E-9 | 1.36E-04±813.4E-9 | 15.62±0.49 | 3.25E-03 |
| 8 | 20.0±5.2 | 6.0±0.16 | 1.60E-03±2.43E-4 | 8.70E-04±8.43E-6 | 26.0±5.36 | 3.57E-03 |
| 15 | 52.0±10.6 | 7.0±0.44 | 3.22E-04±2.48E-5 | 9.75E-04±7.06E-6 | 59.0±11.04 | 3.88E-03 |
| **G20/Annealed Cu** | | | | | | |
| 0 | 3.5±0.70 | 2.0±0.60 | 5.23E-04±2.85E-5 | 5.00E-04±4.02E-6 | 5.5±1.3 | 6.12E-04 |
| 8 | 2.5±0.10 | 1.0±0.03 | 2.99E-04±1.69E-5 | 1.03E-03±8.35E-6 | 3.5±0.13 | 4.04E-04 |
| 15 | 0.5±0.015 | 0.4±0.007 | 9.80E-04±7.22E-5 | 8.46E-04±8.60E-6 | 0.9±0.022 | 1.58E-03 |

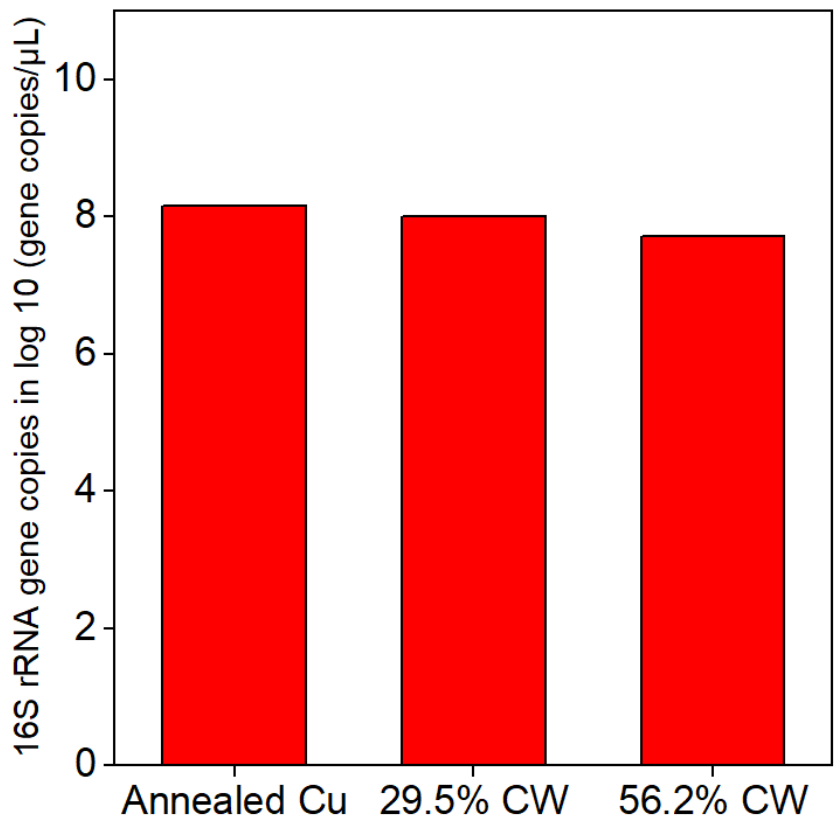

**Figure S2.** Quantification of bacterial biomass from the 16SrRNA gene copies log10 (gene copies / µL)

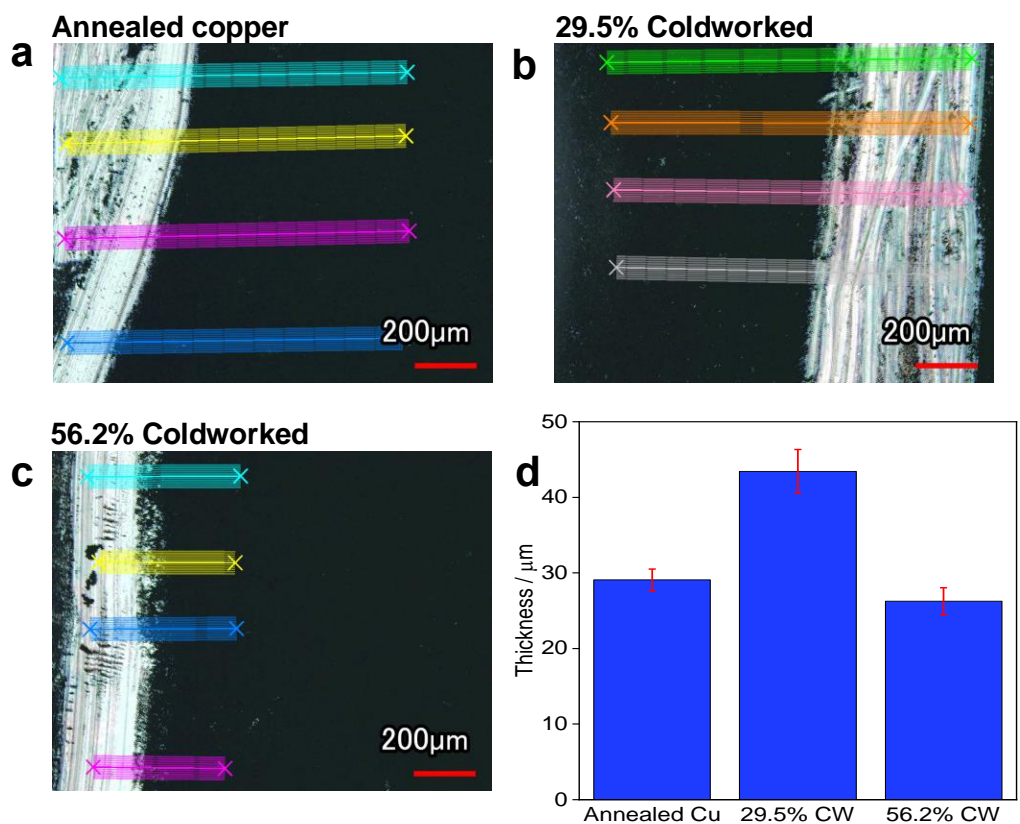

**Figure S3.** CLSM image of dry biofilm after 4 month of 60 days corrosion test- **(a)** annealed Cu, **(b)** 29.5% CW, **(c)** 56.2% CW showing the lines for taking the roughness profile of biofilm (black) and underlying Cu (white) **(d)** average thickness of the biofilm matrix ranging from ~28µm to ~45µm.

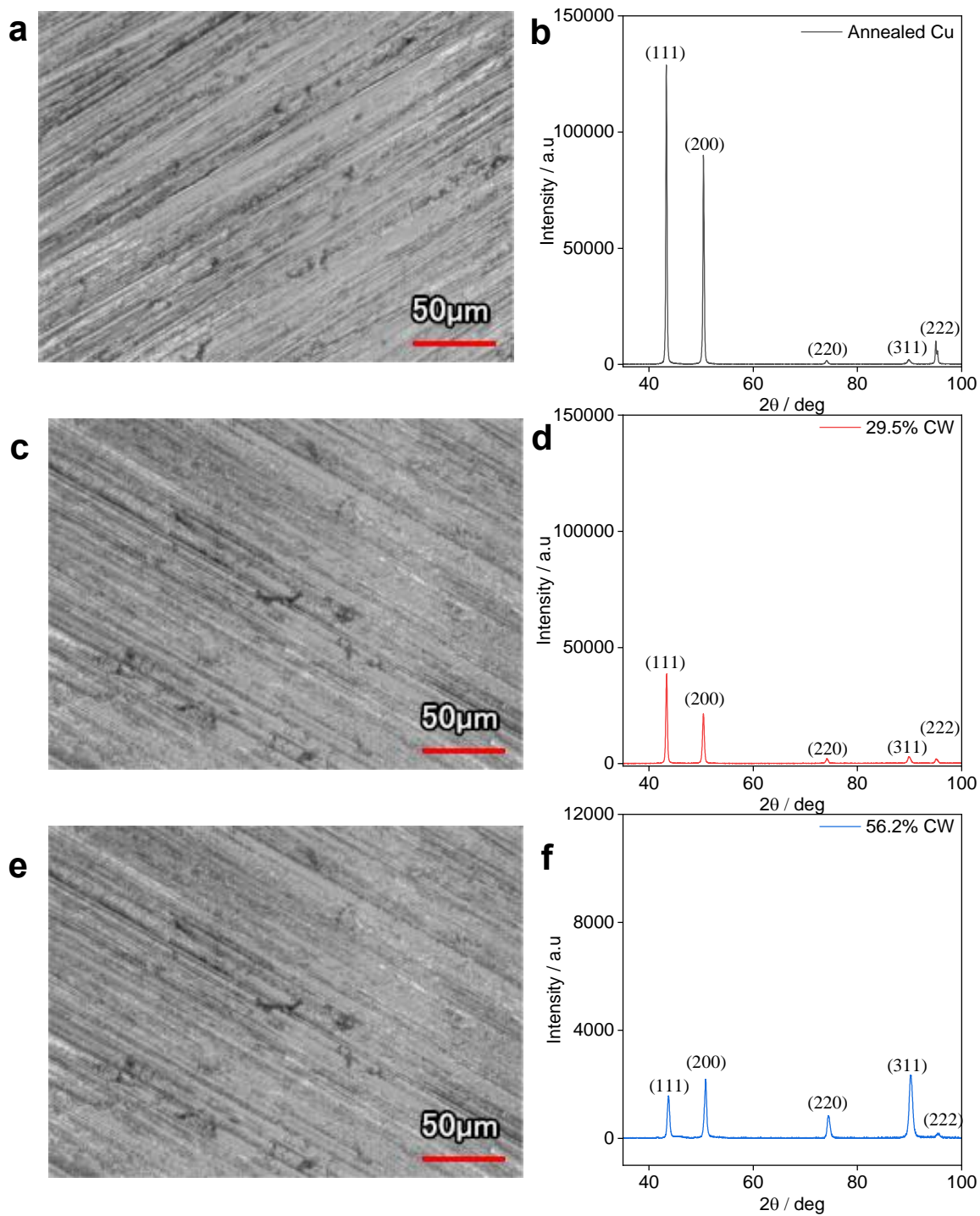

**Figure S4. CLSM image and corresponding XRD for chemical composition before corrosion test- (a, b)** annealed Cu , **(c)(d)** 29.5% CW and **(e)(f)** 56.2% CW showing the absence of biofilm and the presence of polycrystalline Cu with different crystallographic orientation.

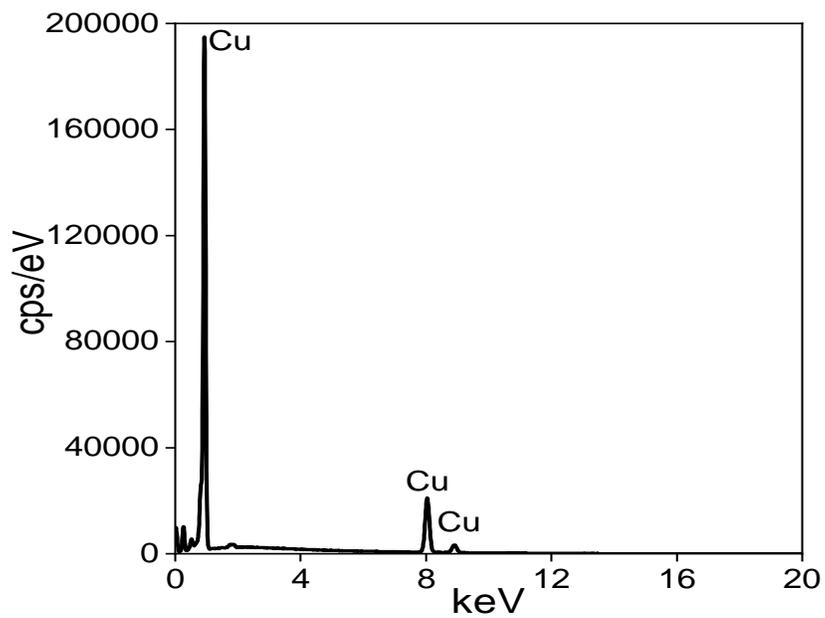

**Figure S5.** EDS analysis- elemental composition of pristine Cu before exposure to co-cultured G20 and MIC21 showing the Cu peaks

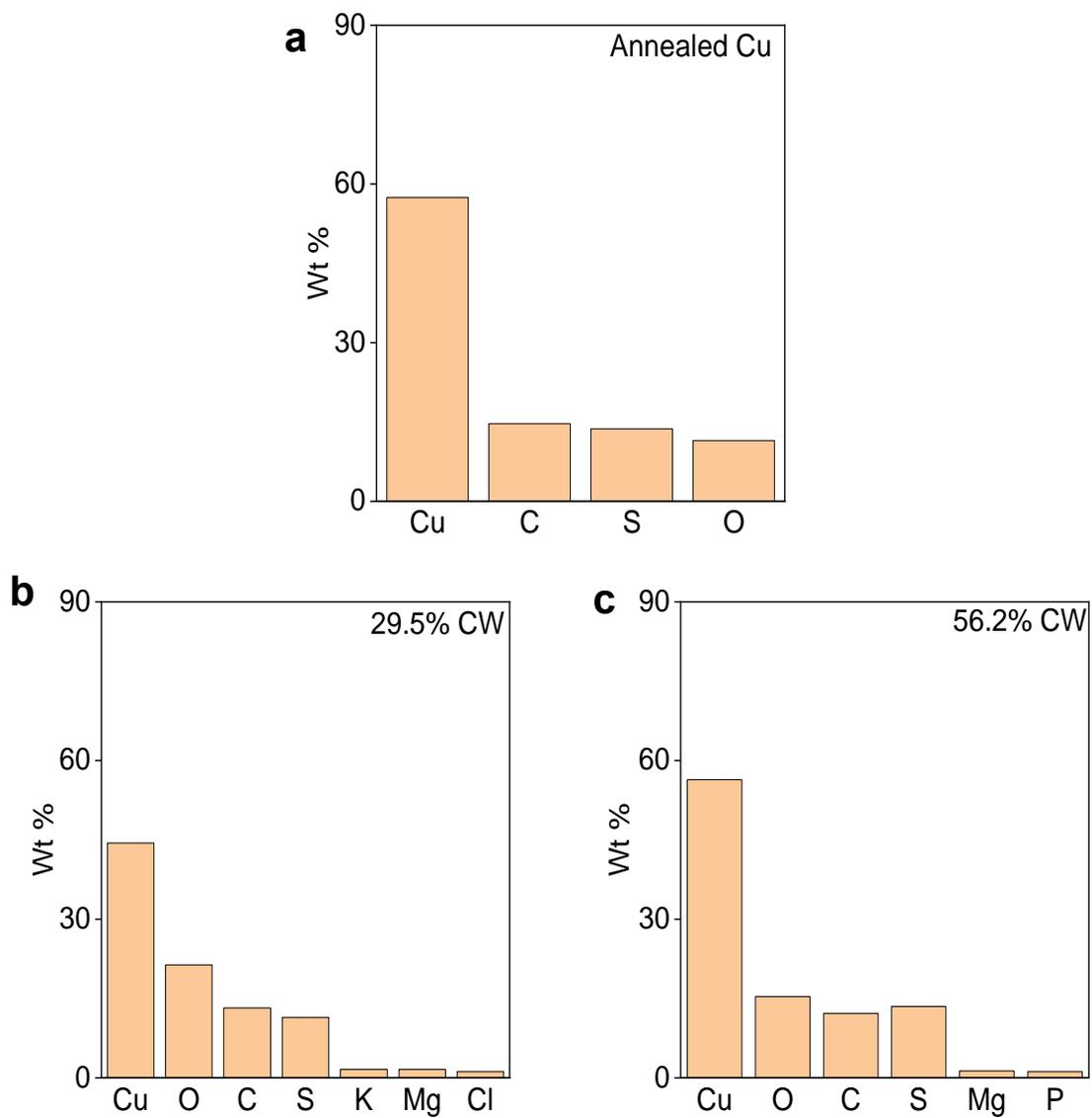

**Figure S6.** EDS analysis- elemental composition of **(a)** annealed Cu **(b)** 29.5% CW and **(c)** 56.2% CW showing the presence of sulfur (S), Oxygen (O) and Carbon (C) as major constituents.

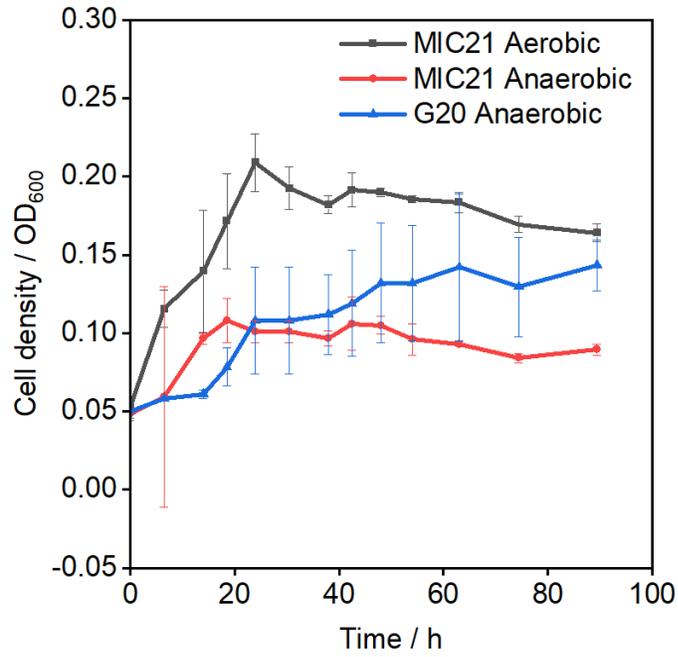

**Figure S7.** Growth Curve for strain MIC21 under aerobic and anaerobic condition and G20 under anaerobic condition

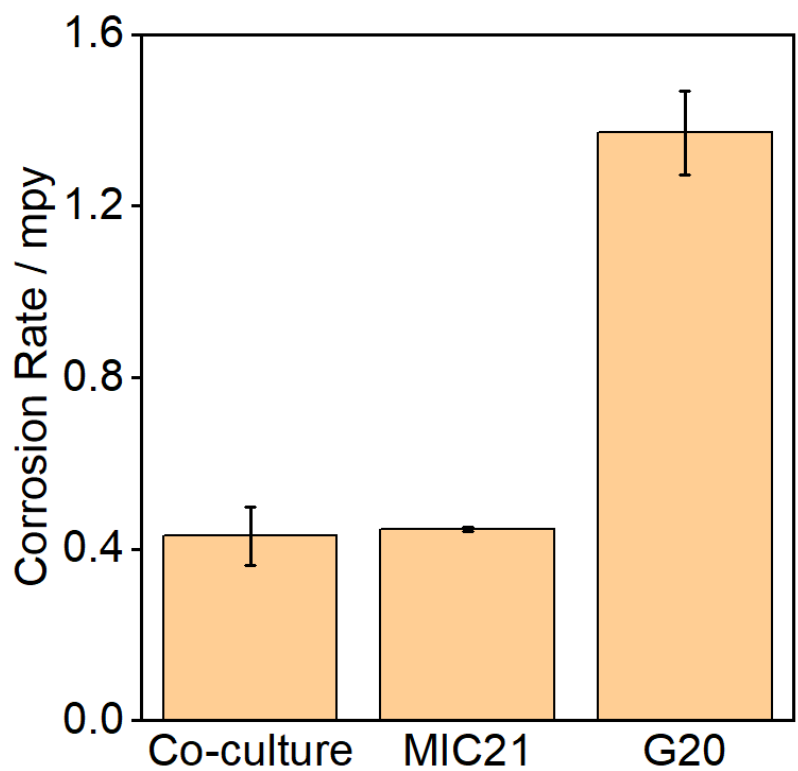

**Figure S8.** Weight loss measurement showed approximately 1-fold less corrosion rate for co-culture when compared to the pure culture of strain MIC21 (0.445±0.005 mpy) and approximately 3-fold less than the pure culture of strain G20 (1.371±0.098 mpy)

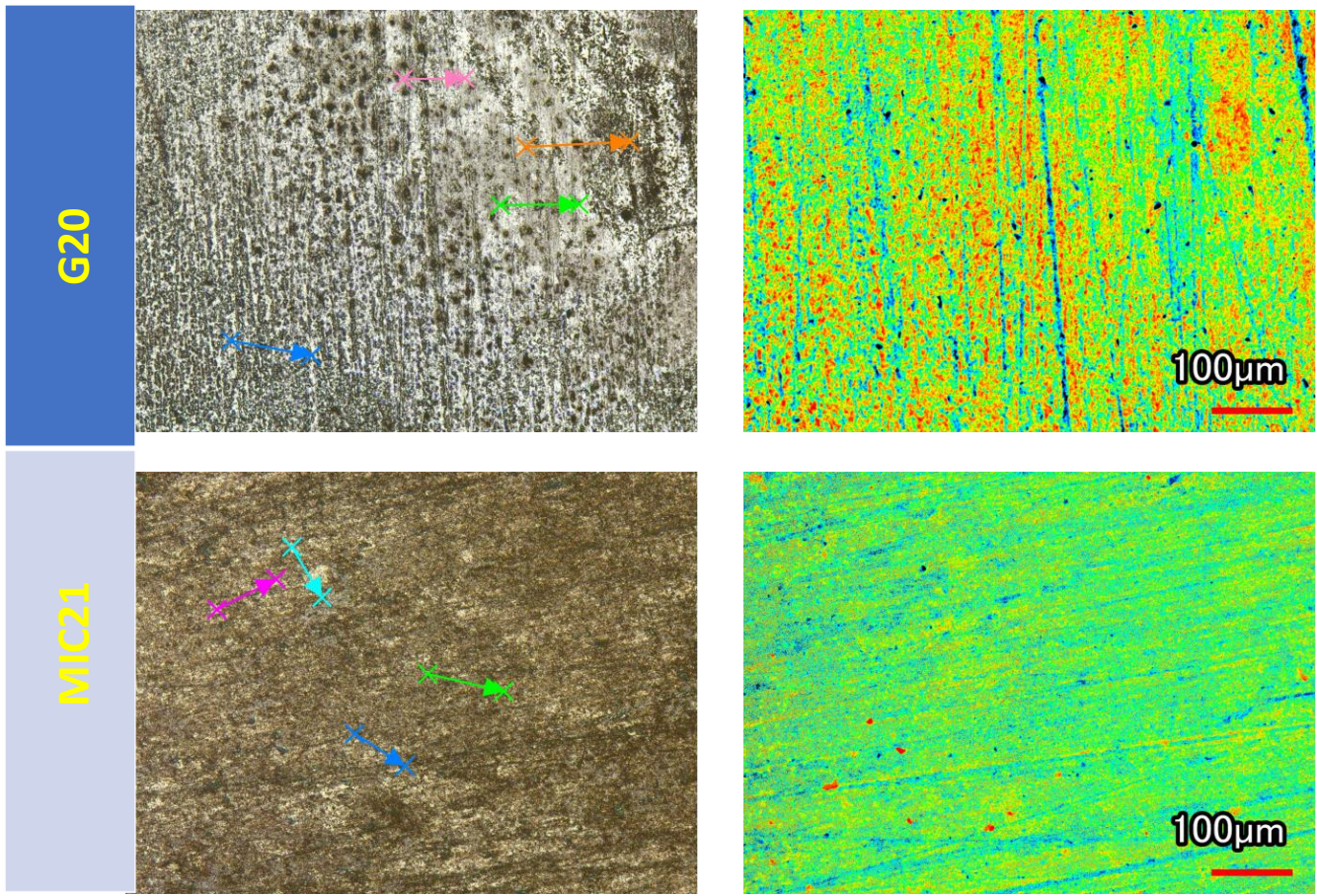

**Figure S9**. CLSM imageses showed multiple pit in annealed Cu exposed to pure culture of *O. alaskensis* strain G20 when compared to *Citrobacter* sp. strain MIC21.

# References

1. Dou, W., et al., *Investigation of the mechanism and characteristics of copper corrosion by sulfate reducing bacteria.* Corrosion Science, 2018. **144**: p. 237-248.
2. Chen, S., P. Wang, and D. Zhang, *Corrosion behavior of copper under biofilm of sulfate-reducing bacteria.* Corrosion Science, 2014. **87**: p. 407-415.
3. Huang, G., K.-Y. Chan, and H.H. Fang, *Microbiologically induced corrosion of 70Cu-30Ni alloy in anaerobic seawater.* Journal of the Electrochemical Society, 2004. **151**(7): p. B434.
4. Chen, S. and D. Zhang, *Study of corrosion behavior of copper in 3.5 wt.% NaCl solution containing extracellular polymeric substances of an aerotolerant sulphate-reducing bacteria.* Corrosion Science, 2018. **136**: p. 275-284.
5. Pedersen, A. and M. Hermansson, *The effects on metal corrosion by Serratia marcescens and a Pseudomonas sp.* Biofouling, 1989. **1**(4): p. 313-322.
6. Ponmariappan, S., S. Maruthamuthu, and R. Palaniappan, *Inhibition of corrosion of mild steel by Staphylococcus sp.* Transactions of the SAEST, 2004. **39**(4): p. 99-108.
7. Guo, Z., et al., *Adhesion of Bacillus subtilis and Pseudoalteromonas lipolytica to steel in a seawater environment and their effects on corrosion.* Colloids and Surfaces B: Biointerfaces, 2017. **157**: p. 157-165.
8. Warthmann, R., et al., *Bacterially induced dolomite precipitation in anoxic culture experiments.* Geology, 2000. **28**(12): p. 1091-1094.
9. Thauer, R.K., K. Jungermann, and K. Decker, *Energy conservation in chemotrophic anaerobic bacteria.* Bacteriological reviews, 1977. **41**(1): p. 100-180.
10. Alasvand Zarasvand, K. and V. Ravishankar Rai, *Identification of the traditional and non-traditional sulfate-reducing bacteria associated with corroded ship hull.* 3 Biotech, 2016. **6**(2): p. 1-8.
11. Tian, F., et al., *Electrochemical corrosion behaviors and mechanism of carbon steel in the presence of acid-producing bacterium Citrobacter farmeri in artificial seawater.* International Biodeterioration & Biodegradation, 2020. **147**: p. 104872.
12. Shahryari, Z., K. Gheisari, and H. Motamedi, *Effect of sulfate reducing Citrobacter sp. strain on the corrosion behavior of API X70 microalloyed pipeline steel.* Materials Chemistry and Physics, 2019. **236**: p. 121799.
13. Zhang, Q., et al., *Corrosion behavior of WC–Co hardmetals in the oil-in-water emulsions containing sulfate reducing Citrobacter sp.* corrosion science, 2015. **94**: p. 48-60.